\documentclass[aip,amsmath,amssymb,superscriptaddress,reprint]{revtex4-2}

\usepackage{charter,graphicx,verbatim,threeparttable,float,amssymb }
\usepackage{setspace}
\usepackage{multirow}
\usepackage[font=small,labelfont=bf,
   justification=Justified,
   format=plain]{caption}
\usepackage{subcaption}
\usepackage{ragged2e}

\usepackage{array}

\newcolumntype{C}[1]{>{\centering\arraybackslash}p{#1}}

\usepackage[version=4]{mhchem}
\begin{document}



\title{Enhancing Corrosion Resistance of Aluminum Alloys Through AI and ML Modeling}

\author{Farnaz Kaboudvand}
\thanks{These three authors contributed equally}
\affiliation{Research Fellow, The Washington Institute for STEM Entrepreneurship and Research, Washington DC, USA}

\author{Maham Khalid}
\thanks{These three authors contributed equally}
\affiliation{Research Fellow, The Washington Institute for STEM Entrepreneurship and Research, Washington DC, USA}

\author{Nydia Assaf}
\thanks{These three authors contributed equally}
\affiliation{Research Fellow, The Washington Institute for STEM Entrepreneurship and Research, Washington DC, USA}

\author{Vardaan Sahgal}
\affiliation{The Washington Institute for STEM Entrepreneurship and Research, Washington DC, USA}

\author{Jon P. Ruffley}
\affiliation{Naval Nuclear Laboratory, West Mifflin, PA 15122, USA}

\author{Brian J. McDermott}
\email{BrianJ.Mcdermott@unnpp.gov}
\affiliation{Naval Nuclear Laboratory, West Mifflin, PA 15122, USA}

\begin{abstract}
Corrosion poses a significant challenge to the performance of aluminum alloys, particularly in marine environments. This study investigates the application of machine learning (ML) algorithms to predict and optimize corrosion resistance, utilizing a comprehensive open-source dataset compiled from various sources. The dataset encompasses corrosion rate data and environmental conditions, preprocessed to standardize units and formats. We explored two different approaches, a direct approach, where the material’s composition and environmental conditions were used as inputs to predict corrosion rates; and an inverse approach, where corrosion rate served as the input to identify suitable material compositions as output. We employed and compared three distinct ML methodologies for forward predictions: Random Forest regression, optimized via grid search; a feed-forward neural network, utilizing ReLU activation and Adam optimization; and Gaussian Process Regression (GPR), implemented with GPyTorch and employing various kernel functions. The Random Forest and neural network models provided predictive capabilities based on elemental compositions and environmental conditions. Notably, Gaussian Process Regression demonstrated superior performance, particularly with hybrid kernel functions. Log-transformed GPR further refined predictions. This study highlights the efficacy of ML, particularly GPR, in predicting corrosion rates and material properties.
\end{abstract}

\maketitle

\section{Introduction}


Aluminum alloys are widely used due to their lightweight properties, high strength, and versatility. However, corrosion resistance remains a significant limitation for their broader adoption. Developing aluminum alloys with both excellent mechanical properties and high corrosion resistance is a critical goal in materials engineering, particularly for applications in aerospace, marine, and structural environments\cite{reboul2011metallurgical, lunder1997corrosion, esquivel2020corrosion}.

Materials design involves two key challenges: the forward problem and the inverse problem. The forward problem aims to establish a clear understanding of how a material’s composition and processing conditions influence its properties. By studying these relationships, researchers can fine-tune alloy compositions and optimize manufacturing parameters to achieve superior performance. On the other hand, the inverse problem takes a different approach, asking: ``What composition and processing conditions are needed to achieve specific material properties?" This is particularly complex for multi-component materials, where the vast compositional space and intricate interdependencies between elements make exploration difficult \cite{juan2021accelerating, raabe2023accelerating}.

Traditional methods to study corrosion resistance, which rely heavily on trial-and-error experimentation, are often time-consuming and resource-intensive. Their iterative nature makes exploring the vast compositional space of multi-component materials inefficient. To overcome these challenges, computational modeling and machine learning (ML) provide efficient alternatives, enabling faster material discovery, optimized compositions, and reduced reliance on costly experimentation \cite{li2020ai, raabe2023accelerating, zhang2020bayesian}.

Artificial intelligence and machine learning have gained significant momentum in materials science for their ability to model and predict material properties, as well as for accelerating materials discovery. In recent years, several methods have been developed that leverage data-driven modeling techniques to identify promising compositions with enhanced corrosion resistance. There has also been an increased emphasis on making materials data publicly available for AI/ML modeling purposes, as evidenced by popular resources such as the Materials Project \cite{Jain2013} and GNoME \cite{merchant2023scaling}.

Recent studies have demonstrated the power of integrating ML with high-fidelity simulations such as ab-initio calculations for alloy design. For instance, Ji et al. \cite{ji2023corrosion} introduced a reinforcement self-learning ML algorithm to predict and optimize stress-strain responses in corrosion-resistant Al alloys. Their computational approach not only identified ideal compositions but also guided the fabrication of Al-Mg-Zn-Cu alloys with experimentally verified performance. Zeng et al. \cite{zeng2024machine} alternatively demonstrated a physics informed method tested on Al-Cr-Fe-Co-Ni high entropy alloy data to identify composition regions leading to higher corrosion protection. Additionally, as experimental data is often limited, there have been several studies focused on effectively utilizing machine learning for smaller materials science datasets as well\cite{smallfeng2019using, xu2023small, ji2022randomsmall}. This, along with other recent research, underscores the practical potential of ML-integrated alloy design strategies for simultaneously achieving enhanced corrosion resistance and superior mechanical properties\cite{ji2023corrosion,zeng2024machine, sasidhar2023enhancing, hu2024designing, tran2020multi, ji2022random}.

The process of corrosion is complex, typically influenced by both the intrinsic properties of the alloy and the environmental conditions in which the material is used. Factors such as elemental composition, temperature, pH, and duration of exposure all contribute to the corrosion behavior\cite{esquivel2020corrosion, harsimran2021overview, riggs1967temperature}. Numerous studies have been conducted to investigate the influence of these variables on the corrosion resistance of aluminum alloys, particularly in chloride-rich environments such as seawater. However, the heterogeneous nature of available experimental data presents challenges for building consistent models.

In this work, we aim to address these challenges by leveraging open-source corrosion datasets, information from literature, converting alloy compositions into atomic percentages, and employing machine learning models to predict corrosion rates across various environments. By evaluating and comparing different ML models —including tree-based methods, feedforward neural networks, and Gaussian process regression — we propose optimized workflows for corrosion rate prediction and alloy performance enhancement.

\section{Methodology}

\subsection{Data Collection and Preprocessing} 
The data for this study was sourced mainly from open repositories such as \textit{NIST CORR-DATA} \cite{Ricker1997}, \textit{Materials Project} \cite{Jain2013}, and \textit{GNoME} \cite{merchant2023scaling}, as well as review articles\cite{MG-ATRENS2020989}, books \cite{Knovel_Aluminum_Database, Vargel2020} and public reports \cite{osti_1069242}. 

Most corrosion data available was recorded for commercially produced alloys, and sources included various attributes including alloy processing conditions, exposure environments, and observed or measured corrosion rates. Alloy identifiers such as the Unified Numbering System (UNS) for Metals and Alloys were used to extract elemental composition data, sourced through the Aluminum Alloy Database \cite{Knovel_Aluminum_Database}, MatWeb \cite{MatWeb2025}, \textit{aluminum.org}\cite{Aluminum2025}, \textit{copper.org}\cite{Copper2025}, and other similar public resources. Ideal composition levels were noted in weight percentages, and subsequently converted to atomic percentages, in order to allow for further computational analysis when necessitated. 

Corrosion damage metrics had been recorded as either corrosion rates -- listed in exact measurements such as mils per year or millimeters per year -- or as grades such as ``A" for excellent resistance to corrosion, up to ``D" for minimal resistance to corrosion. Utilizing supplementary information provided with the \textit{CORR-DATA} database, these grades were converted to approximate numerical values in both mils per year and millimeters per year, signifying exactly the corrosion rates and extent of corrosion damage\cite{Ricker1997}.

This data was then shortlisted to limit the dataset to certain environments more specific to seawater corrosion, such as plain water, seawater, sodium chloride solutions, and other relevant environments. The finalized dataset consisted of 331 samples for corrosion measurements of various different aluminum, magnesium, copper, and other alloys. Features included 32 alloying elements in atomic percentage, 9 different environments recorded as categorical variables, temperature measured in degrees Celsius, and corrosion exposure duration in days as main descriptors. Compositional and environmental information was present for all data samples, while there were $164$ data samples with recorded temperature, $187$ with recorded duration, and $115$ samples with both latter features present. The dataset is summarized in Table \ref{tab:datasumm}, and numerical specifications of alloy compositions are described in Table \ref{tab:ELEMENTS}. Corrosion rate measurements were recorded both in mils per year and millimeters per year, though we utilize mils per year due to its larger presence in corrosion literature. 

\begin{table}[t]
    \centering
    \caption{Summary of variables included in final dataset, and their counts. The final dataset includes values for 32 different alloying elements and 9 environmental categories for all 331 samples. Temperature and exposure duration were recorded for smaller subsets of the data.}
    \begin{tabular}{|c|c|}
        \hline
        \textbf{Variable} & \textbf{Number of Samples} \\
        \hline
         
        Elements (Count: 32)  &  331\\
        Environments (Count: 9)  &  331\\
        Temperature (deg C) &  164\\
        Duration (days) &  187 \\
        Corrosion Rate (mils/year) & 331 \\
        \hline
    \end{tabular}

    \label{tab:datasumm}
\end{table}

\begin{table*}
\centering
\caption{Summary of elements' atomic percentage values in the dataset, and count of alloys containing each element.}
\begin{tabular} {C{2.5cm} C{2.5cm} C{2.5cm} C{2.5cm} C{2.5cm} C{2.5cm} C{2.5cm}}
\hline

\textbf{Element} & \textbf{Count} & \textbf{Max} & \textbf{Element} & \textbf{Count} & \textbf{Max} \\
\hline
Al & 180 & 99.996 & Mg & 97 & 99.959 \\
Si & 168 & 7.187 & Zn & 218 & 99.994 \\
Li & 5 & 9.007 & Ti & 88 & 1.308 \\
Ni & 134 & 71.769 & Cu & 265 & 99.918 \\
As & 4 & 0.033 & Au & 2 & 0.014 \\
B & 2 & 0.063 & C & 82 & 1.184 \\
Ca & 16 & 0.322 & Cd & 33 & 0.001 \\
Co & 24 & 2.654 & Ga & 13 & 0.006 \\
Hf & 13 & 2.335 & In & 2 & 0.030 \\
Mo & 44 & 10.433 & Nb & 23 & 2.486 \\
O & 20 & 1.042 & Pb & 123 & 0.093 \\
P & 32 & 0.082 & S & 56 & 0.087 \\
Sn & 44 & 2.029 & Th & 2 & 0.353 \\
V & 19 & 0.430 & W & 7 & 1.276 \\
Zr & 21 & 96.402 & Fe & 289 & 98.923 \\
Mn & 232 & 1.583 & Cr & 108 & 35.336 \\

\hline
\end{tabular}
\label{tab:ELEMENTS}
\end{table*}

\subsection{Machine Learning Algorithms}
We investigated the utility of various methods, including random forests, ensemble models with gradient boosting, neural networks, and Gaussian process regression. Due to the small size and imbalanced nature of the dataset, we prioritized using methods that would improve predictive accuracy while also preventing overfitting and non-generalizability. We proceed to go over each model employed and general architecture. 

\begin{figure}[H]
    \centering
    \includegraphics[width=0.8\linewidth]{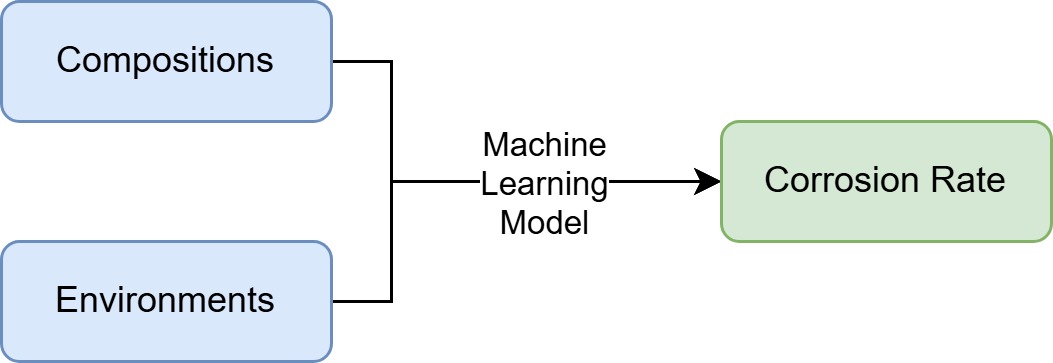}
    \caption{Forward Model Basic Workflow}
    \label{fig:forward}
\end{figure}

\subsubsection{Random Forest Regression}

To model the non-linear relationships present in our experimental corrosion dataset, we employed a Random Forest Regressor using the scikit-learn\cite{sklearn_api} implementation. Random forests are an ensemble learning method that aggregate predictions from multiple decision trees, enhancing predictive performance and accuracy, along with mitigating overfitting — an important consideration given the multi-dimensional nature of our dataset\cite{ISL2013}. 

In our study, the random forest model was used to predict corrosion rates from both compositional and environmental inputs. We applied grid search for hyperparameter tuning, specifically optimizing the number of estimators, tree depth, and feature subset size for each split. Feature importance analysis was also performed, allowing us to identify and prioritize the most influential alloying elements and environmental variables.

This approach proved effective at capturing complex interactions in the data while remaining computationally efficient. Its ability to handle missing values and accommodate heterogeneous input features made it particularly well-suited for our dataset, which integrates information from multiple sources with varying levels of completeness.

\subsubsection{Neural Network}
Neural networks are extremely useful for complex regression and classification tasks, and have been widely applied in practice in several fields, including materials science\cite{takamoto2022towards, witman2023defect, DNN_FENG2019300}. Additionally, as universal function approximators \cite{Approximator}, neural networks are capable of sufficiently approximating any given continuous function given a large enough network architecture, and this property holds despite the complexity of the data it attempts to model.  

In this paper, due to limitations on the amount of available data, and small dataset size, we chose to work with a fully connected deep neural network (DNN), foregoing some of the more complex and state-of-the-art model structures. These deep neural networks have fairly straightforward architectures consisting of an input layer, a variable number of hidden layers, activation functions and finally an output layer. Each hidden layer applies an affine transformation followed by a non-linear activation function to propagate the data forward in between layers. These layer-wise transformations can be simply summarized as 
$$h^{(l)} = g^{(l)} \big(W^{(l)\top} h^{(l-1)} + b^{(l)}\big)$$ where $h^{(l)}$ is the $l^{th}$ hidden layer, $W$ is a matrix of weights, $b$ is a vector of biases, and $g$ is the applied activation function \cite{Goodfellow-et-al-2016}. We can represent the input vector $x$ as $h^{(0)}$. Once the data is propagated through the network fully, the loss, or error is calculated through some specified loss function. The error is subsequently backpropagated, and the weights and biases are then updated based on minimization of the loss through gradient based optimization techniques. 

We utilized the PyTorch\cite{pytorch} framework to build our models, and customized the DNN by testing various network widths -- number of neurons in each layer, depths -- number of hidden layers, activation functions which introduce non-linearities in the models, and applicable loss and optimization functions. Throughout our experiments, we used the Rectified Linear Unit activation function, defined as $$\text{ReLU}(x) = \max (0, x)$$ as this penalizes any negative outputs, and the Huber Loss function\cite{Huberhastie2009elements}, defined as $$
L_\delta(y, \hat{y}) = 
\begin{cases}
\frac{1}{2}(y - \hat{y})^2 & \text{if } |y - \hat{y}| \leq \delta \\
\delta \cdot \left( |y - \hat{y}| - \frac{1}{2} \delta \right) & \text{otherwise}
\end{cases}
,$$ where $\delta$ is a threshold parameter that can be chosen. Adam (Adaptive Moment Estimation) was utilized as our optimization function. Hyperparameters were then tuned through trial and error. Methods to prevent overfitting and improve learning such as dropout regularization and initialization were tested as well, though did not benefit the models. 

\subsubsection{Gaussian Process Regression}
Gaussian Process Regression (GPR) has various applications within the machine learning research community, given its adaptiveness to non-linear and complex data, and its ability to inherently quantify uncertainty. Since GPR is also a non-parametric model, this further allows more flexibility in the model. This is because the parameter space is then infinite and not fixed, unlike in traditional nonlinear regression methods, which tend to model underlying functions through a specified learned function \cite{GP_wang2023}. Due to the nature of the data present, we selected GPR as one of our comparative modeling frameworks, and evaluated its effectiveness for our problem. 

A Gaussian process itself is a stochastic process, and this probabilistic nature is what allows the model to have more accurate representations of complex data. This is extended to a Bayesian, non-parametric regression framework that has high adaptability for various problems. The regression framework itself is modeled by a multivariate Gaussian distribution and can be described as
$$P(f|\textbf{X}) = \mathcal{N}(f | \mu, \textbf{K}),$$ where $\textbf{X}$ defines the observed data points, $f$ is a vector of respective function values, $\mu$ defines the mean functions, and $\textbf{K}$ is the kernel, or covariance function. A Gaussian process can be fully specified through the mean and kernel functions \cite{gaussian_williams2006}, and while the mean function can often be chosen as zero or a constant value, the choice of covariance functions largely influence the effectiveness and modeling properties of one's model. 

A covariance function is a positive semi-definite function that generally defines the similarity between two different function evaluations. Given two input vectors $\mathbf{x, x'}$, the covariance function can be written as $$\text{cov}(f(x), f(x')) = k(x, x')$$ and generally specifies if the given functions are likely under the GP. There are several possible kernel functions, and the choice is largely problem dependent. In this paper, we employed the squared exponential kernel, or the radial basis function (RBF) $$k_{RBF}(x, x') = \sigma^2 \exp\bigg({-\frac{(x - x')}{2 \ell ^2}}\bigg)$$
and the Matérn kernel\cite{Materngenton2001classes}
$$k_{\text{M}}(x, x') = \frac{2^{1 - \nu}}{\Gamma(\nu)} \left( \frac{\sqrt{2\nu} \, |x - x'|}{\ell} \right)^\nu K_\nu\left( \frac{\sqrt{2\nu} \, |x - x'|}{\ell} \right)
$$
along with combinations of the two thereof. Here, $\ell$ defines the length-scale, $\sigma^2$ the variance, $\Gamma$ the gamma function, $K_\nu$ a modified Bessel function, and $\nu$ a positive parameter, whose values are often taken to be $\frac{1}{2}, \frac{3}{2}$ or $\frac{5}{2}$. 

Length-scales specify the smoothness of the kernel functions, and are often learned or optimized through the Gaussian process algorithm, with smaller length-scales implying more rapid changes in function behavior. While the RBF kernel is still fairly smooth even with different length-scales due to its infinite differentiability, the $\nu$ parameter in the Matérn kernel modifies the kernel's smoothness in a more direct manner. Smaller values of $\nu$ lead to less differentiability in the function, and therefore a higher ability to model discontinuous regions. According to Stein\cite{stein1999interpolation}, this property of Matérn kernels may in turn lead to more realistic modeling of physical phenomena that inherently have non-smooth behaviors. Notably, the RBF kernel is in fact a special case of the Matérn kernel; as $\nu \to \infty$, the Matérn kernel converges to the RBF kernel\cite{stein1999interpolation}. 

Combining kernels is another approach that has been researched widely\cite{additive_hastie2017generalized}. Adding kernels together can often allow for extrapolation, leading to modeling data far from available training inputs effectively\cite{duvenaud2014automatic}. As the sum of two covariance functions is also positive semi-definite, we know this is still a valid kernel for the Gaussian process model.

Additionally, given that corrosion rate measurements are inherently positive quantities, we must also take into account the positivity constraints in our model, in order to ensure both the underlying assumptions and subsequent predictions are more accurate. There are several methods to deal with data such as this, including using transformations, warping functions \cite{GPsnelson2003warped}, and non-Gaussian likelihood functions \cite{GPswiler2020survey}. We chose a simple, yet historically effective method, i.e., the log-transformed Gaussian Process, wherein we transformed the output variables through a logarithm, modeled the data through a regular Gaussian process, and then transformed the predicted variables back to their original scales. 

The models were implemented through the GPyTorch \cite{gpytorch} framework, which allows for more flexibility and customization during the model construction process. For more details on the general theory of Gaussian Processes, we refer the reader to the standard text \cite{gaussian_williams2006}.

\subsubsection{Ensemble Method}
Ensemble models are larger meta-learner models that take the predictions of two or more base models and average out or aggregate their predictions in order to increase model accuracy. This is quite useful in the case of data with missing values, as separate models can be trained for each subset of data that contains certain features, which may not be present throughout the dataset, and then the results are compiled in a weighted fashion based on specified accuracy scores to produce better estimates. 

While one of the models used was a random forest regressor, the other was a gradient boosting method. Gradient boosting differs slightly from a random forest, by virtue of the fact that it builds the decision trees sequentially, and rather than fitting on the output data directly, it is fitted on the residuals or errors instead\cite{ISL2013}. We used the scikit-learn\cite{sklearn_api} library for the random forest model, and XGBoost \cite{xgboost} to build the gradient boosting model.

\begin{figure}
    \centering
    \includegraphics[width=0.8\linewidth]{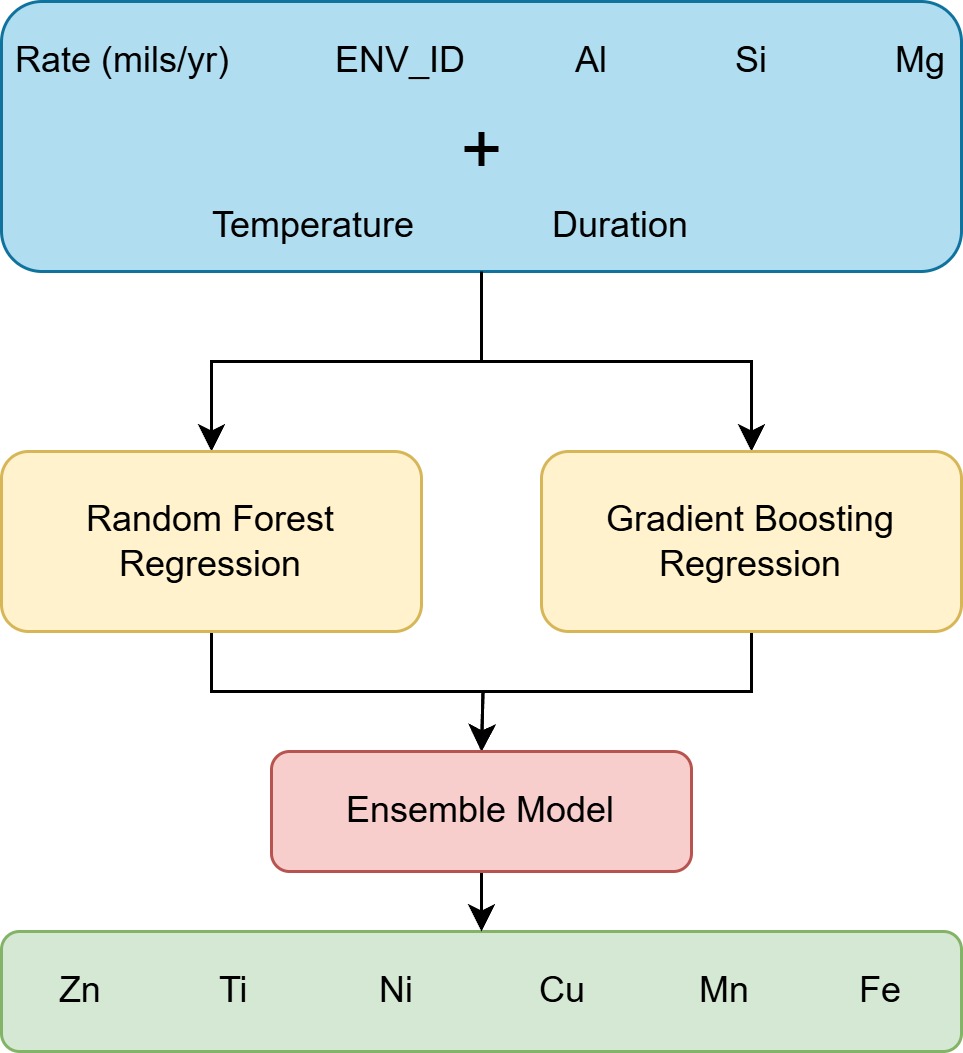}
    \caption{Inverse Regressor Base Model Workflow}
    \label{fig:inverse}
\end{figure}

For our experiments, an initial base model contained the features ``Corrosion Rate", ``Environment" variables numerically encoded as ``Environment ID", and atomic percentages of the elements Al, Si, and Mg. Additional subsets of data used included corrosive environment exposure duration -- containing 187 samples -- as well as both duration and exposure temperature values -- with 115 total samples. This created a total of three different weighted submodels. As these were multi-targeted regression models, other elements' atomic percentages were predicted through the specified inputs, though we utilized a limited subset based on some elements of interest, as well as those with the most available non-zero entries. Repeated samples' predictions for those present in all three submodels were aggregated and averaged, and the final model was based on the predictions of all submodels. A high-level overview of the workflow, along with input and output variables used can be seen in Fig. \ref{fig:inverse}.

\subsection{Model Evaluation}
The performance of each model was assessed using metrics such as the coefficient of determination (R$^2$) score, mean absolute error (MAE) and mean squared error (MSE). Each of these metrics help in assessing understand how well a given model works. 

An R$^2$ score describes how well the variance in a model's dependent or target variable can be explained by the regression model and independent variables. A score that is closer to 1 indicates a better model fit, showing that a high proportion of the variance in the model can be described by the input variables, while a score closer to 0 indicates worse fit. 

MAE and MSE are used to measure how far off each predicted value is from the true experimental values, and while similar, both have their own advantages over each other. MAE calculates the absolute difference between the observed values $y_i$ and their respective predictions $\hat{y_i}$, and averages it out over all samples. MSE on the other hand, squares the distances between each pair before taking the average. Root Mean Squared Error (RMSE) is the square root of the MSE, and is utilized in order to understand the errors within the context and scale of the actual values, making it more interpretable. Therefore, an ideal model would work to minimize average errors while maximizing the calculated R$^2$ score.

\section{Results and Discussion}

\subsection{Forward Modeling}

\begin{figure}
\begin{subfigure}{0.5\textwidth}
    \centering
    \includegraphics[width=\linewidth]{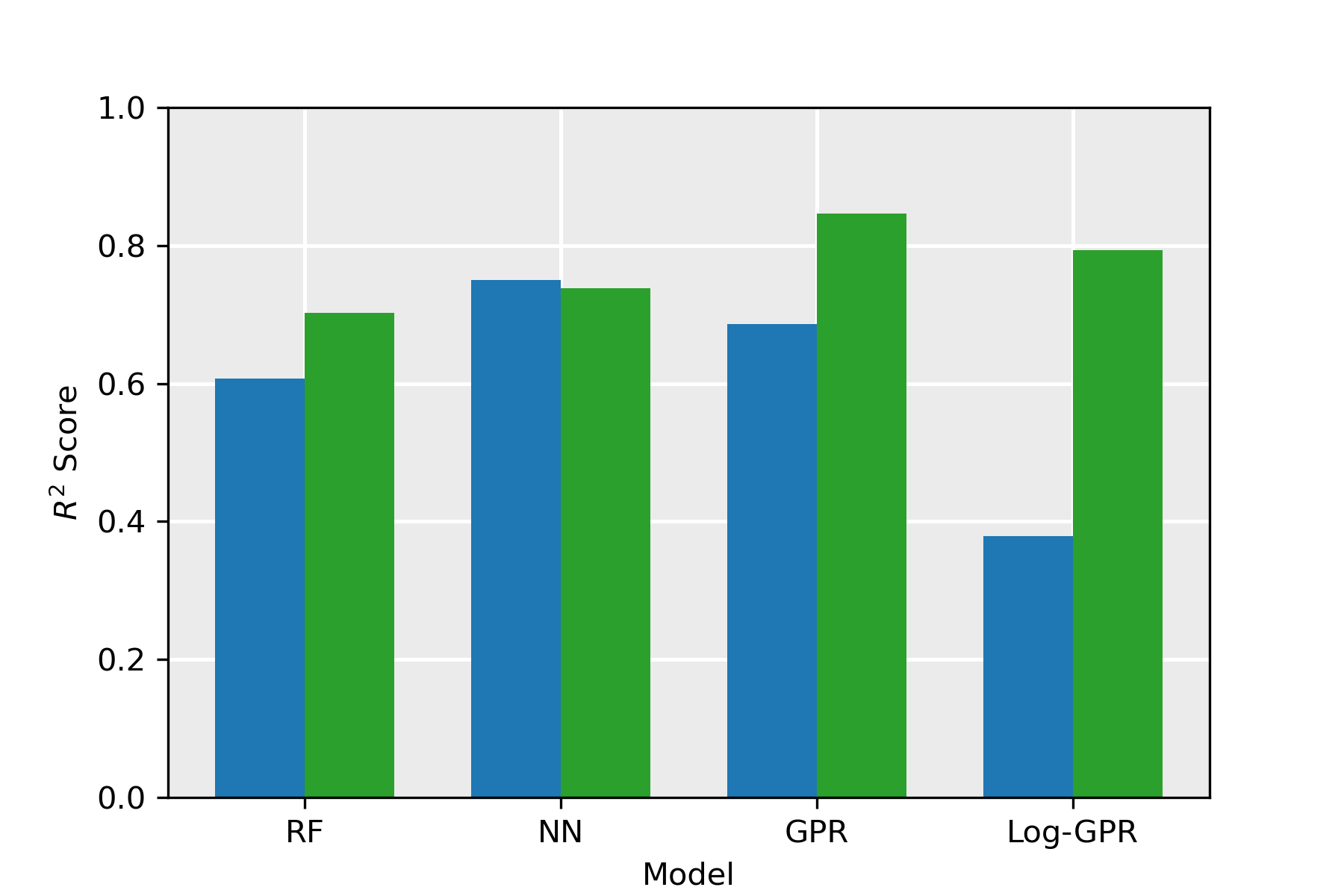}
    \caption{R2 Scores for All Models}
    \label{subfig:R2}
\end{subfigure}
\begin{subfigure}{0.5\textwidth}
    \centering
    \includegraphics[width=\linewidth]{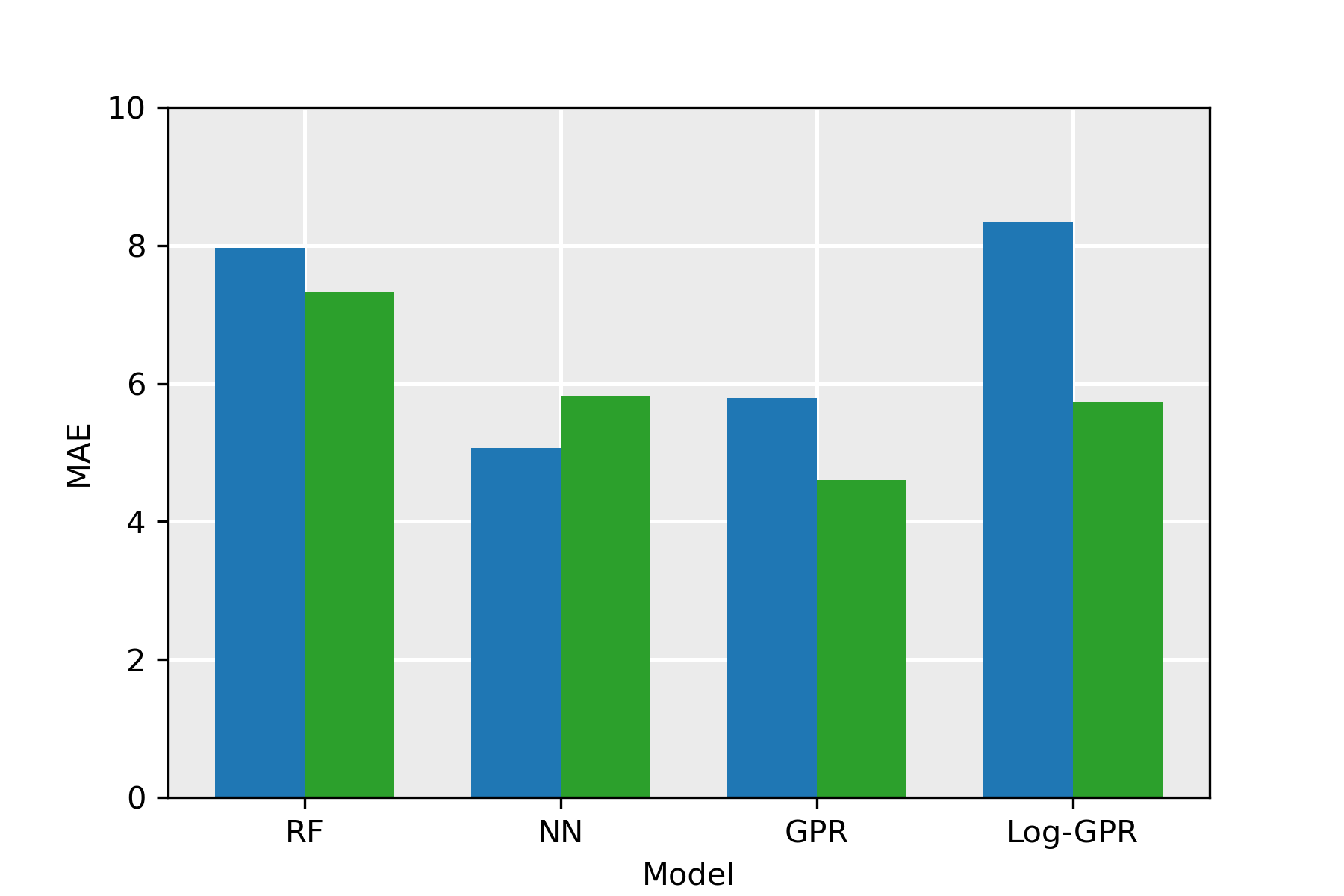}
    \caption{MAE for All Models}
    \label{subfig:MAE}
\end{subfigure}
\begin{subfigure}{0.5\textwidth}
    \centering
    \includegraphics[width=\linewidth]{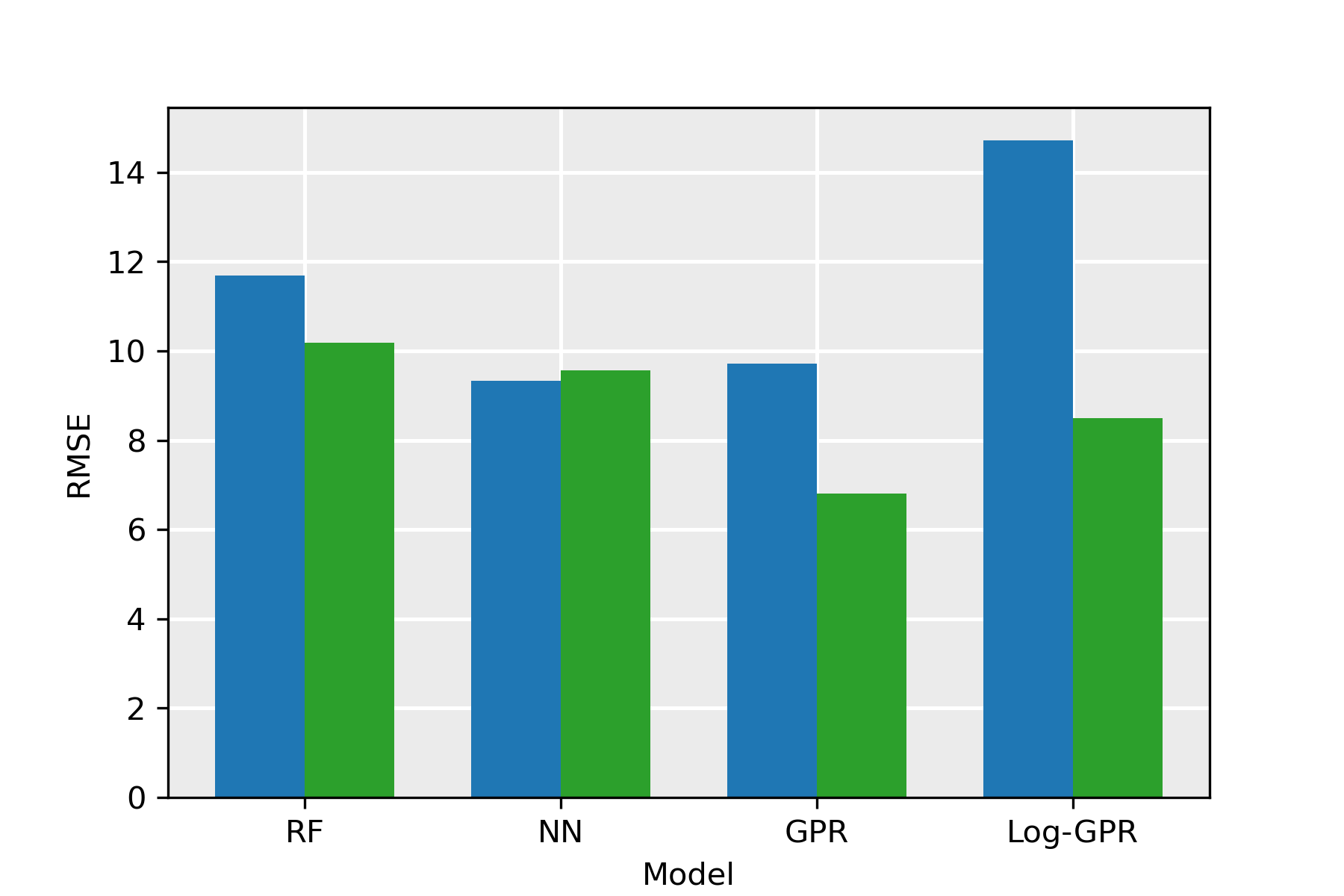}
    \caption{RMSE for All Models}
    \label{subfig:RMSE}
  \end{subfigure}
  \caption{Comparison of performances of different models to predict corrosion rates. Blue bars signify Composition-Only models, while green bars show Composition+Environment models. Here, \ref{subfig:R2} compares the respective $R^2$ scores of each model, \ref{subfig:MAE} compares the Mean Absolute Error, and \ref{subfig:RMSE} compares the Root Mean Squared Error for each model.}
  \label{fig:Comparison Results}
\end{figure}

Throughout forward modeling experiments, we utilized all elemental composition information, as well as environmental information as we compared the performance of each model with these features (Fig. \ref{fig:Comparison Results}). As the environments variable is a categorical feature, for these models, we used one-of-$k$, or one-hot encoding, in order to include each category as separate binary feature columns, with a 1 indicating presence of that environment, or a 0 indicating the absence. For the DNN and GPR models, we also applied a standard scaler in order to set the mean of the data to $0$ and the variance to $1$. 

\begin{figure*}
    \begin{subfigure}[b]{0.94\textwidth}
        \centering
        \caption{Random Forest Regression}
        \includegraphics[width=0.46\textwidth]{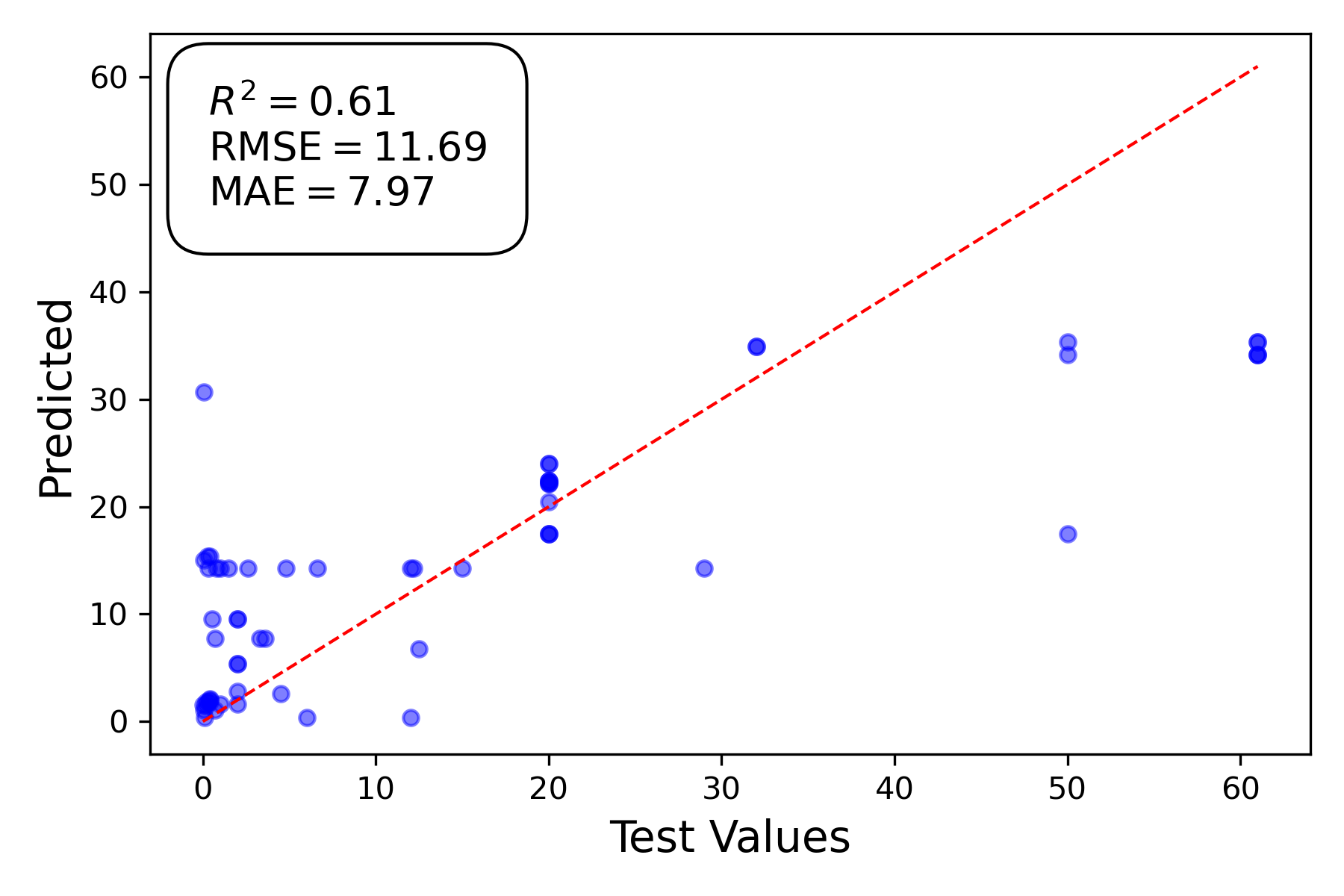}
        \includegraphics[width=0.46\textwidth]{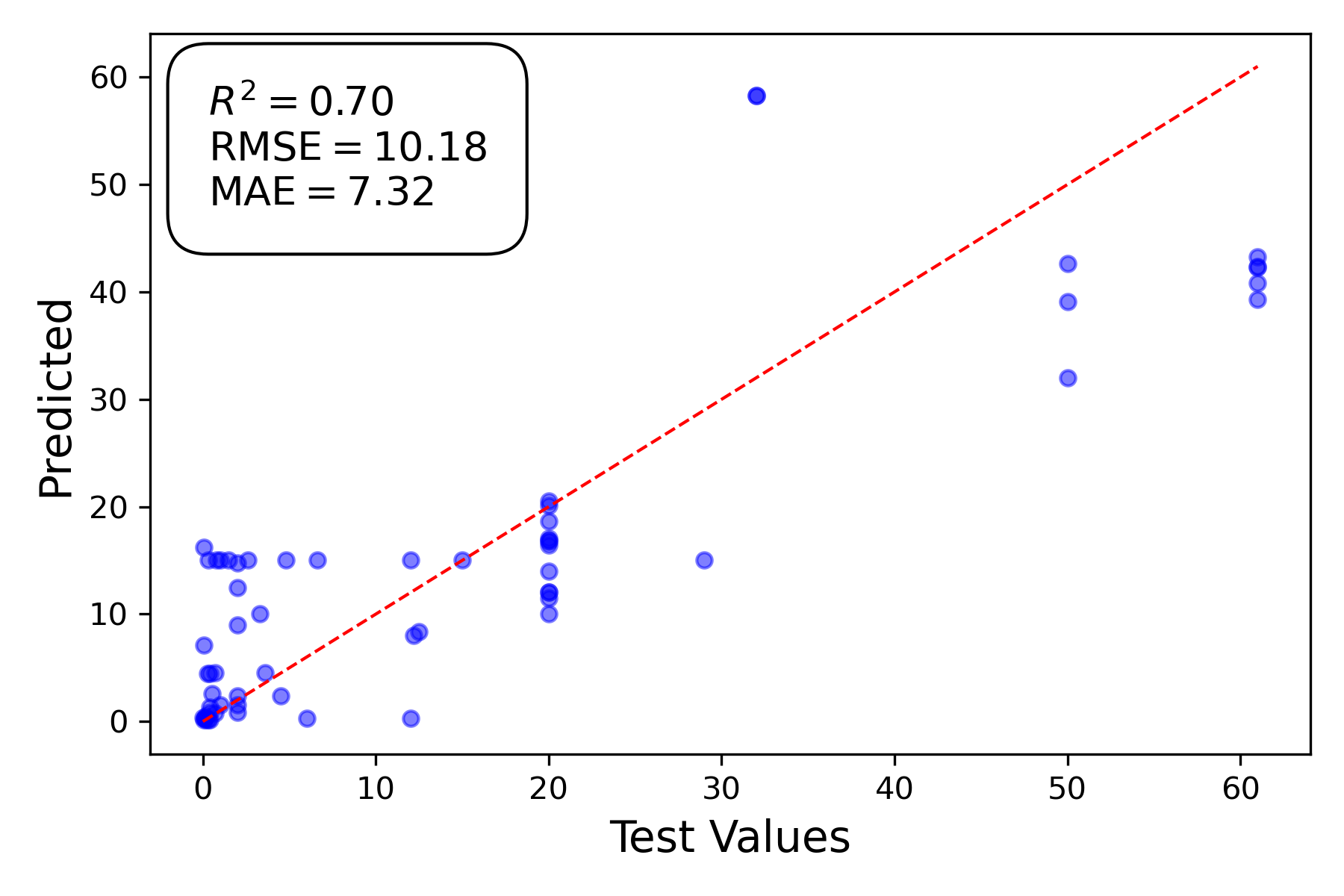}
        \label{subfig:RF}
    \end{subfigure}
    \begin{subfigure}[b]{0.94\textwidth}
        \centering
        \caption{Deep Neural Network}
        \includegraphics[width=0.46\textwidth]{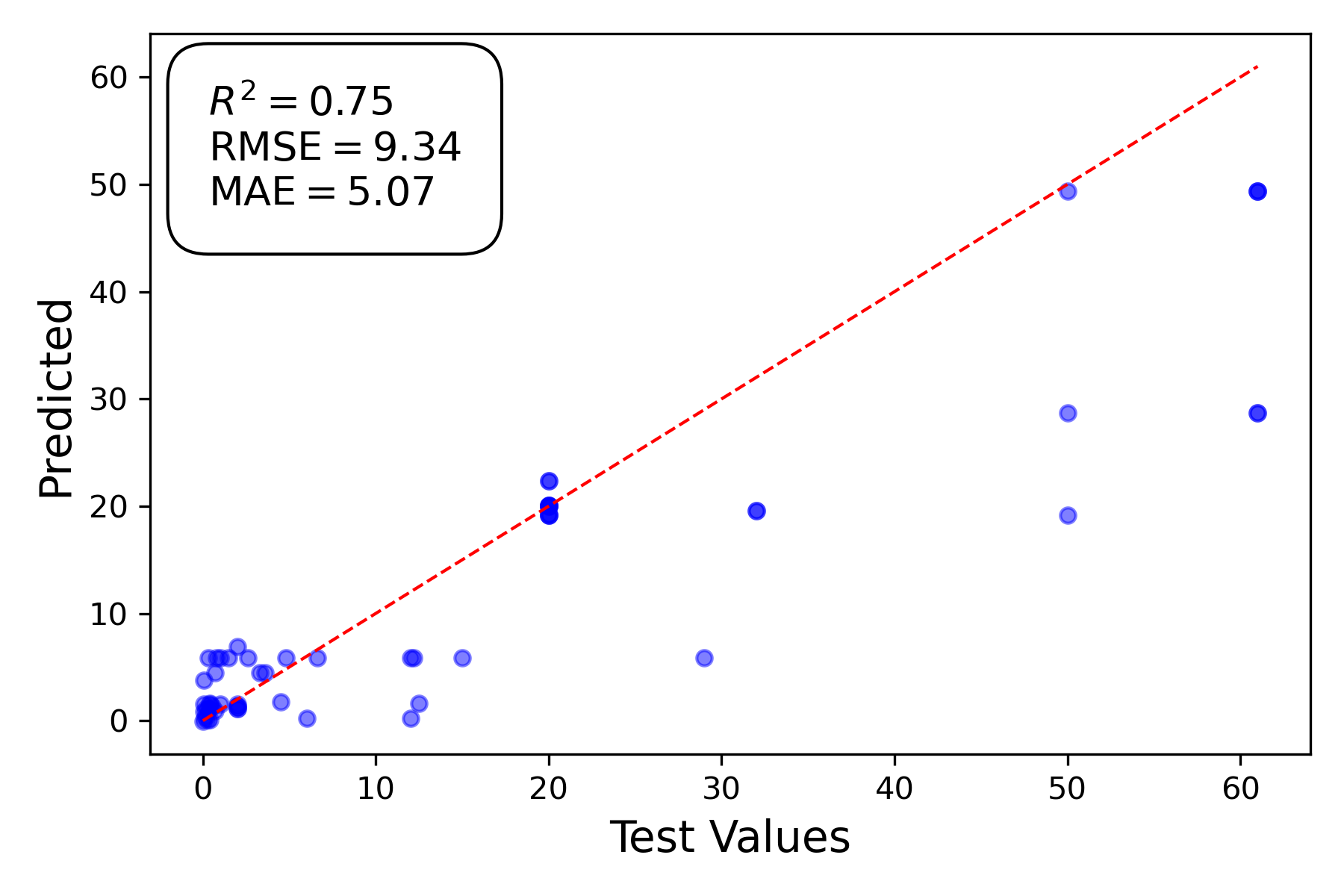}
        \includegraphics[width=0.46\textwidth]{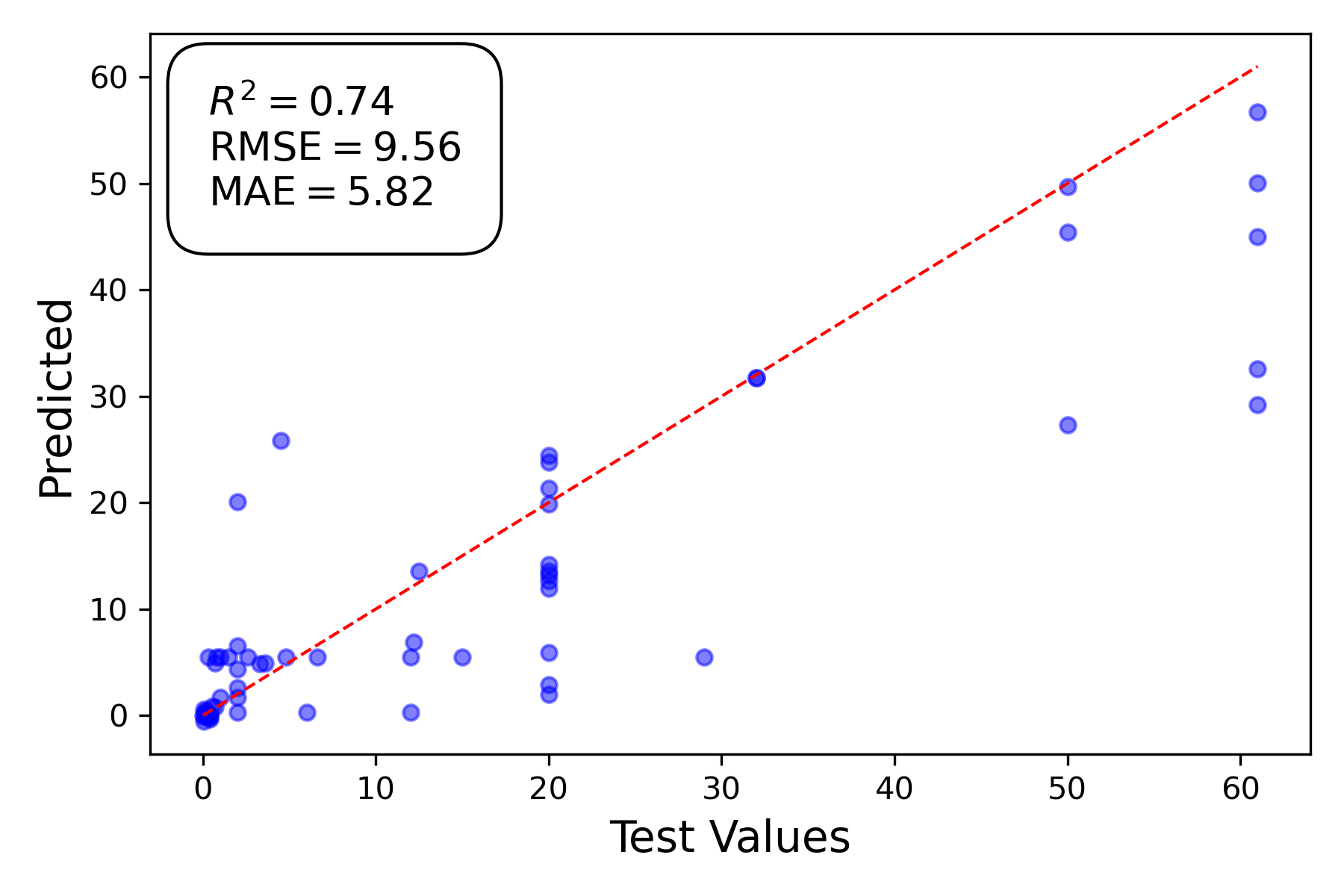}
        \label{subfig:NN}
    \end{subfigure}
    \begin{subfigure}[b]{0.94\textwidth}
        \centering
        \caption{Gaussian Process Regression}
        \includegraphics[width=0.46\textwidth]{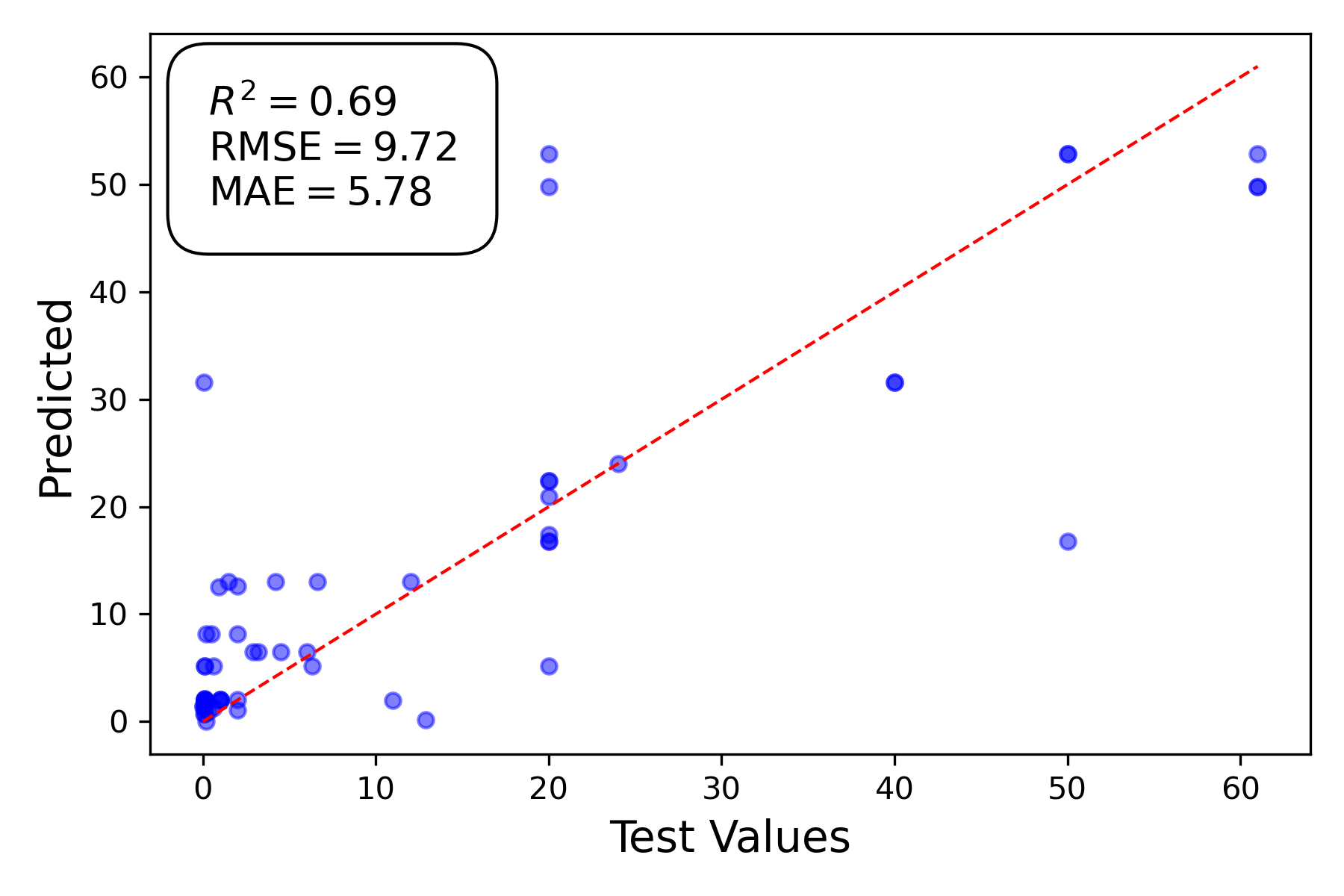}
        \includegraphics[width=0.46\textwidth]{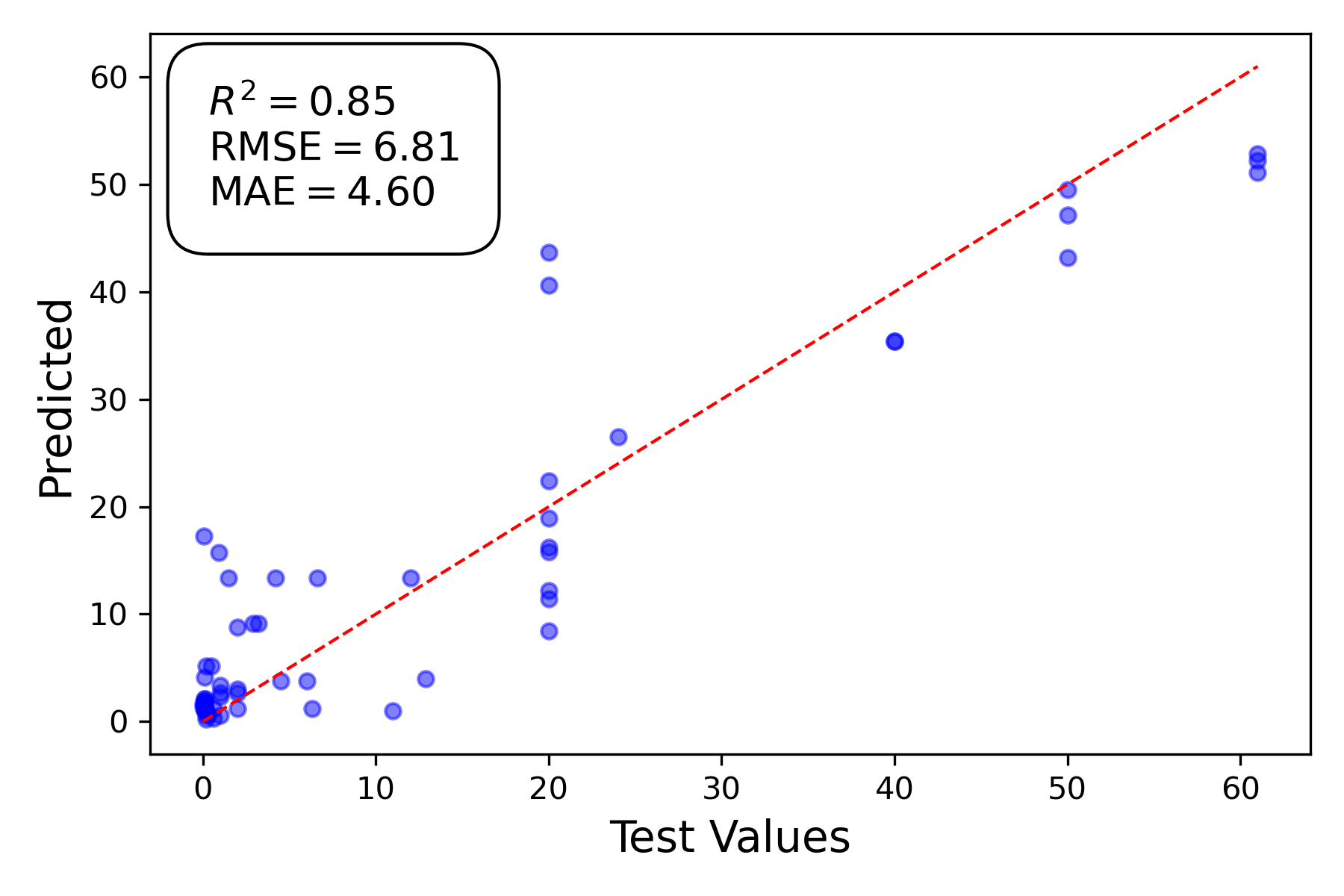}
        
        \label{subfig:gprcm32}
    \end{subfigure}
    \begin{subfigure}[b]{0.94\textwidth}
        \centering
        \caption{Log-Transformed Gaussian Process}
        \includegraphics[width=0.46\textwidth]{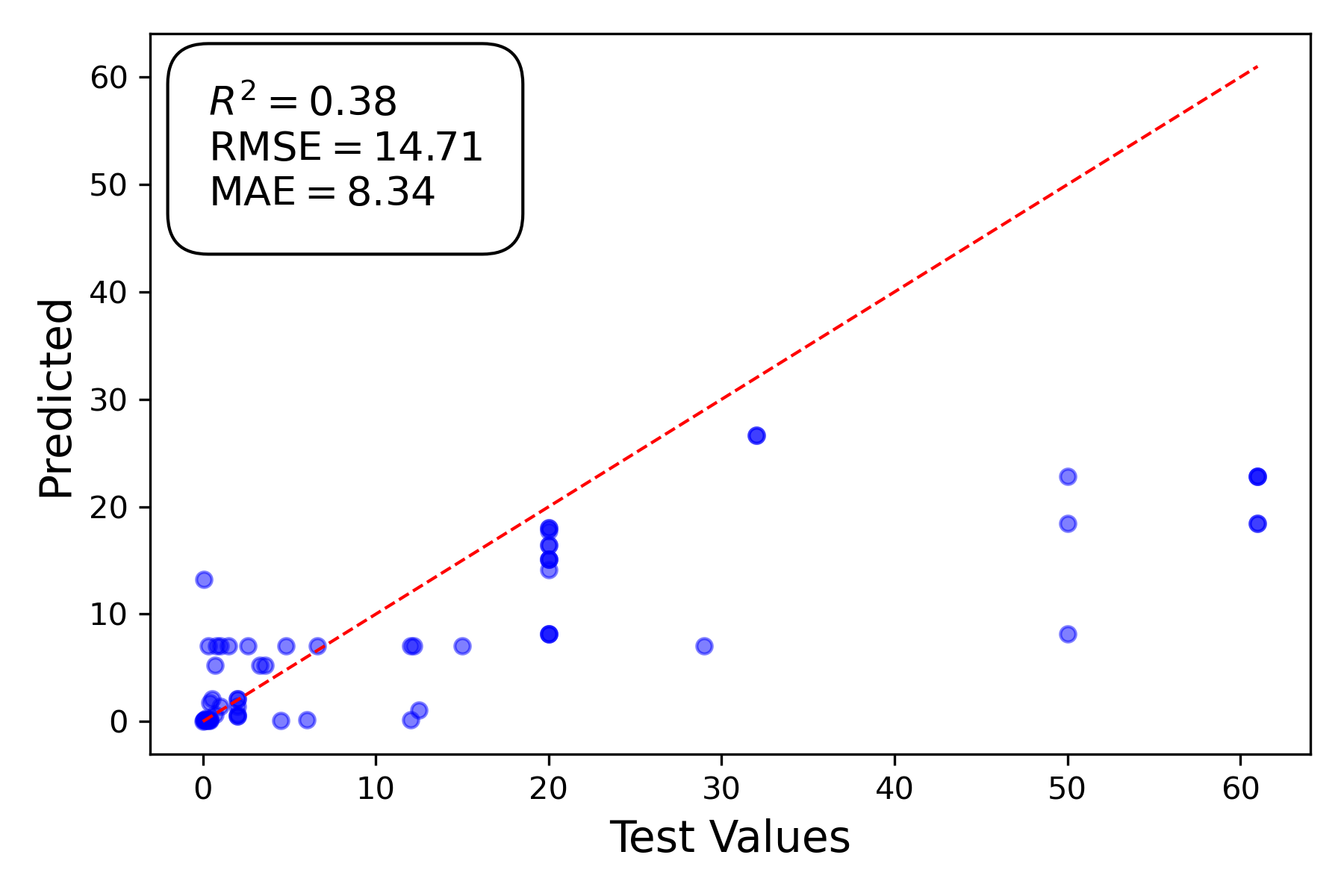}
        \includegraphics[width=0.46\textwidth]{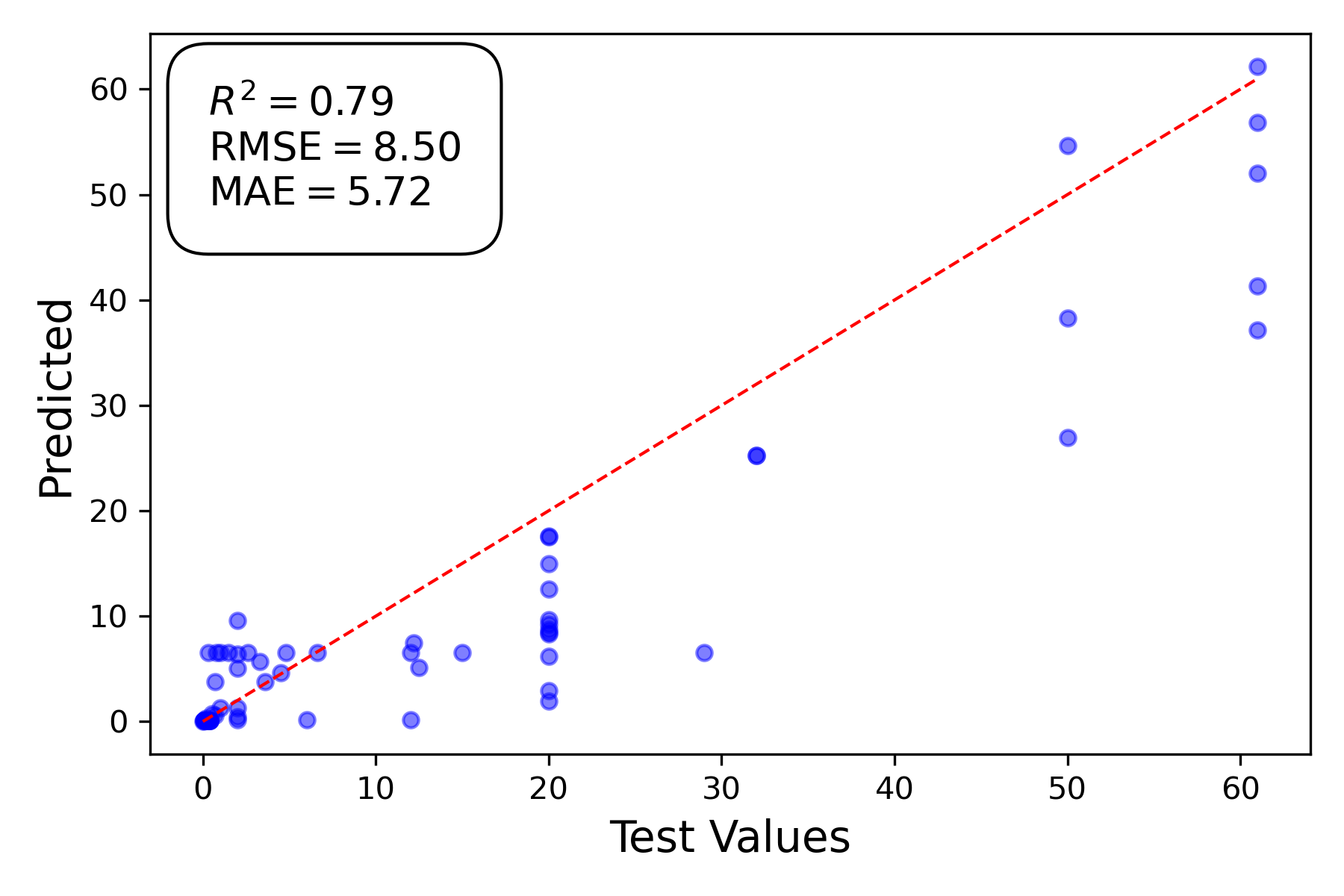}
        \label{subfig:lgprc}
    \end{subfigure}
    \captionsetup{justification = Justified, font=small, labelfont=bf, singlelinecheck=off}
    \caption{Performance comparison between four modeling frameworks and two feature sets. Each row shows the modeling frameworks as described. The left four subfigures use atomic compositions only, and the right four subfigures use additional environmental features as well. Each plot displays the true vs. predicted corrosion rates.}
    \label{fig:allmodels}
\end{figure*}

All forward models were trained using a randomized $80-20$ train-test split, with output data capped at $100$ mils per year, in order to remove the sparse outliers. We first tested models using only compositional data, to understand to what extent only elements included in an alloy affected corrosion rates. We then tested while also utilizing the environments each alloy was exposed to, and compared the predictive results in Fig. \ref{fig:allmodels}. 

The random forest regressor was built through the \textit{scikit-learn} \cite{sklearn_api} library in \textit{Python}. After choosing an initial set of hyperparameters, we utilized a grid search method in order to choose the best set of hyperparameters across various subsets of the data through 5-fold cross validation. This model was tested for the composition-only features, as well as the composition and environment dataset. We found that the predictive accuracy measured through three specified metrics was at a relatively lower level, though increased with the addition of more informational features (Fig. \ref{subfig:RF}). 

The deep neural network model showed a comparatively better ability to approximate corrosion rate measurements as compared to the random forest regressor.  The models were built with an input layer, four hidden layers of sizes $64, 32, 16$ and $8$, and an output layer of size $1$. Each layer was followed by a ReLU activation function. While some loss functions such as MSE loss were tested, we ultimately found the best performance with a Huber loss function, while setting the $\delta$ parameter to $0.1$. As there is a large variation in the corrosion rate measurements, ranging from scales of $10^{-4}$ to $10^2$, the $\delta$ value was selected in order to prevent excessive sensitivity to outliers or large errors. The optimization function used throughout our experiments was Adam, with a learning rate of $0.001$. Both composition and composition plus environment models (Fig. \ref{subfig:NN}) were run for $200$ epochs, stopping as the loss values plateaued in order to prevent overfitting. While dropout regularization and weight initialization were explored, we found these did not benefit the models. In contrast to what we saw with the other forward models, we found that for the deep neural network, the addition of environment features did not significantly affect the accuracy metrics, and in fact reduced the R$^2$ score while increasing both error metrics. 

We also explored the efficacy of Gaussian process regression (GPR) frameworks for this problem. The regression fit is highly dependent on the kernel function chosen, as variations within each tend to significanty affect the results. An important component of these kernel functions is also the length-scale parameter. While most readily available kernel function implementations do not do so by default when fitting a model, we utilized automatic relevance determination (ARD) -- a method to modify length-scales, fitting them separately for each feature rather than the same value throughout. As we have input data of varying scales, along with one-hot encoded variables for models including categorical variables, ARD was necessary in order to fit the length-scale independently \cite{duvenaud2014automatic}, based on each feature separately rather than all features combined. 

A Gaussian likelihood function and constant mean function were used for all GPR models. Additionally, we utilized the Adam optimization function through the \textit{PyTorch} framework, and the number of training epochs was varied, being reduced in the cases of ill-conditioned matrices or convergence issues as training progressed. A number of kernel functions were tested, along with combinations thereof, and we found that the Matérn kernel's inherent ability to capture non-linear relationships greatly benefitted the models' predictions. For the straightforward GPR model, a Matérn kernel with the parameter $\nu$ set to a value of $1.5$ produced the best results, as seen by the overall RMSE, MAE and R$^2$ scores.  

We observed that especially in comparison to other models, the GPR was able to not only capture lower values of corrosion rates where data was abundant, but higher corrosion rates as well (Fig. \ref{subfig:gprcm32}). This demonstrated the ability of Gaussian processes to capture various complex relationships in the data more effectively as compared to other models. 

For the log-transformed Gaussian process, or log-GPR as we call it, we utilized the same setup as the regular GPR, however with one major difference -- the output variables were transformed through a natural logarithm before being scaled. Due to the nature of this transformation function, while predictions can be highly accurate within the log-transformed space, one must take care to prevent large errors within this space, as the inverse transform introduces an exponentiated, multiplicative error, rather than an additive one. Large errors and deviations from the values within the log space are therefore substantially amplified when reverted back into the original space. 

We tested several kernel functions, and found that the utilization of a sum of RBF and Matérn kernel, with $\nu = 2.5$, along with the inclusion of the categorical environment variable, produced more accurate results, though underestimated some corrosion rate values to a greater extent than some other models (Fig. \ref{subfig:lgprc}). 

While not available for all data samples, we also explored the effectiveness of including information such as exposure temperature or duration, and observed that for our forward models, while duration of exposure did not have a significant effect on model accuracy, temperatures seemed to influence the evaluation metrics more, often increasing R$^2$ scores up to $0.89$ with the log-GPR model. These were, however, largely reduced datasets as compared to the original, raising questions about model generalizability to more data. Nonetheless, the observed temperature effect aligns with prior studies on the relationship between corrosion behavior and temperature \cite{kritzer2004corrosion}.

\subsection{Inverse Modeling}
We also investigated an inverse, multi-targeted regression approach, similar to experiments by Bhat et al \cite{bhat2024inverse}. We utilized a significantly reduced input space, and a multidimensional output space containing elements of interest that may positively or negatively affect corrosion rates, as summarized in Fig. \ref{fig:inverse}. Additionally, the categorical environment feature was encoded numerically, assigning a single number to each environment listed. These factors make the inverse model markedly different from the forward models. 

\begin{figure}[H]
    \begin{subfigure}{0.45\textwidth}
        \centering
        \includegraphics[width=\linewidth]{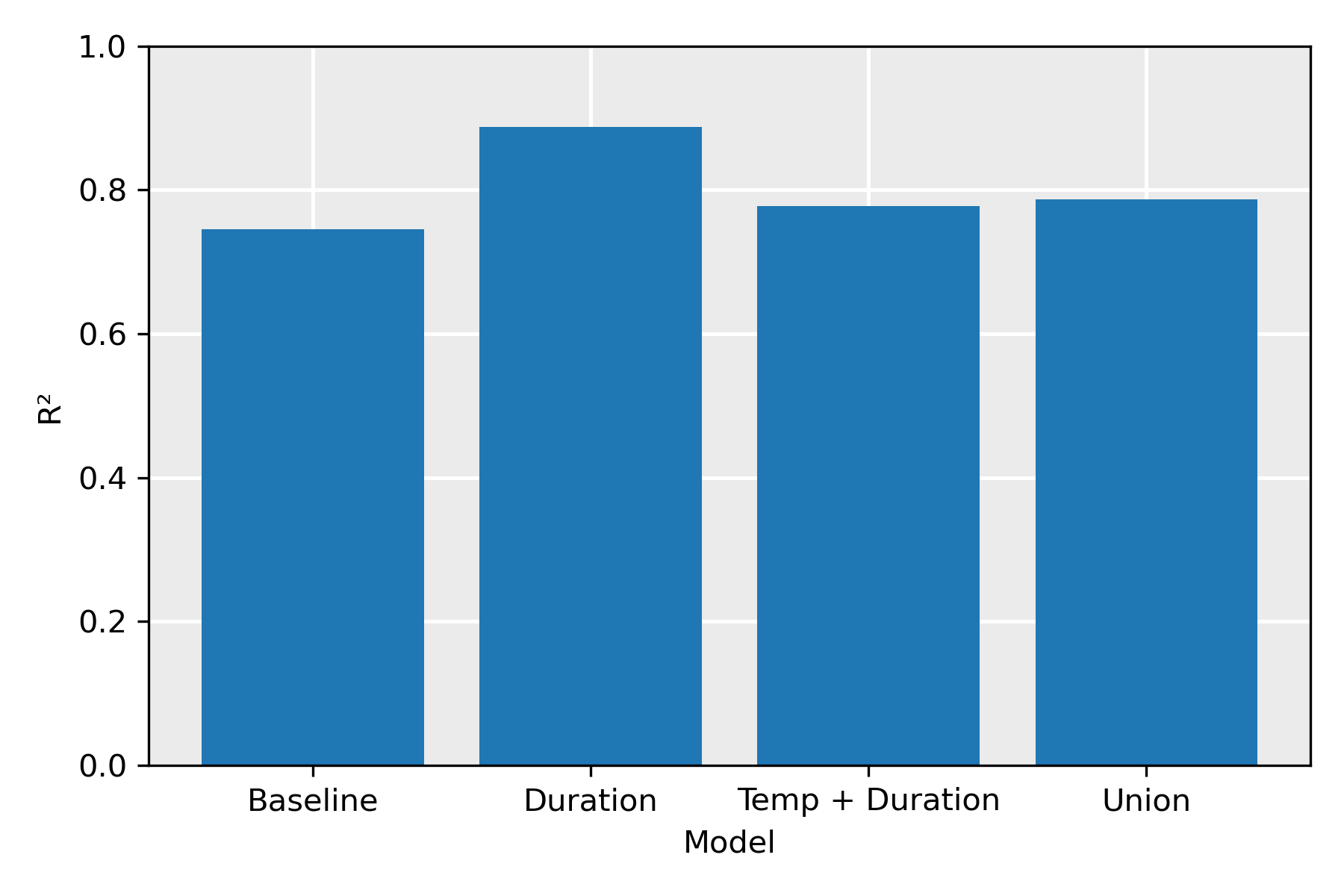}
        \caption{R2 Scores}
        \label{subfig:Nydr2}
    \end{subfigure}
    \begin{subfigure}{0.45\textwidth}
        \centering
        \includegraphics[width=\linewidth]{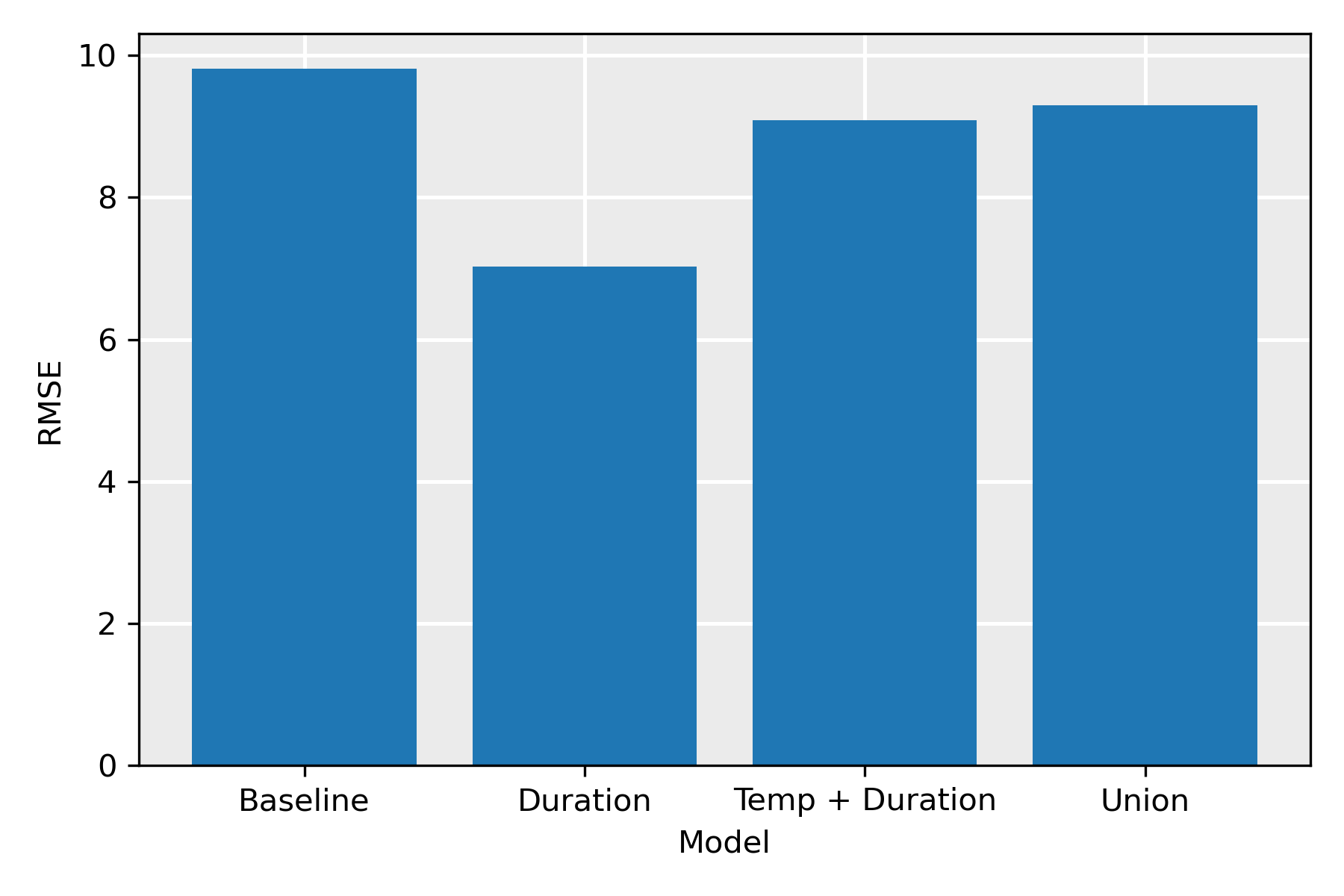}
        \caption{MSE}
        \label{subfig:NydMSE}
    \end{subfigure}
    \captionsetup{justification = justified, font=small, labelfont=bf}
    \caption{Accuracy Metrics for Inverse Model, where Fig. \ref{subfig:Nydr2} compares $R^2$ scores when including different features, while Fig. \ref{subfig:NydMSE} compares the respective RMSE scores. Baseline contains the corrosion rate, environment, and atomic percentages of the elements Al, Si, Mg. Each subsequent bar shows which other features are added to the baseline.}
  \label{fig:Inverse Comparison Results}
\end{figure}

Hyperparameters such as the number of estimators or maximum depth were tuned for both the random forest and gradient boosting models, and a learning rate of $0.1$ was chosen for the gradient boosting regressor. The multi-output regressor was fitted on both models, and results aggregated, utilizing three different sets of input features. Each submodel as an ensemble of the two different regression models was evaluated separately, and then also aggregated for jointly present samples. We found that the accuracy of the submodels, as well as the aggregated model, was quite high, as can be seen in Fig. \ref{fig:Inverse Comparison Results}. The best performing inverse regressor overall was the model including the base set of features described in Fig. \ref{fig:inverse} -- Corrosion Rate, Numerically Encoded Environment variables, and atomic percentages for Al, Si, and Mg -- as well as duration information, and was trained on the subset in which duration values were present, which resulted in an R$^2$ score of $0.89$ and an RMSE value of $7.03$. 

\begin{table}[H]
\centering
\caption{R\textsuperscript{2} and RMSE scores for different elements' predicted atomic percentages through the ensemble model with scores aggregated from 3 different sub-ensemble models.}
\begin{tabular}{|C{2cm} | C{2cm} | C{2cm}|} 
\hline
\textbf{Element} & \textbf{R$^2$} & \textbf{RMSE} \\
\hline
Zn & 0.91 & 8.762 \\
Ti & 0.87 & 0.099 \\
Ni & 0.86 & 8.411 \\
Cu & 0.85 & 13.485 \\
Fe & 0.56 & 13.761 \\
Mn & 0.68 & 0.267 \\
\hline
\end{tabular}
\label{tab:element_scores}
\end{table}

Individual R$^2$ scores and RMSE values for each output value in the union model are listed in table \ref{tab:element_scores} as well, in which we can see the efficacy of using a small amount of features to predict a large number of possible element compositions based on corrosion rates and main alloying elements. This union model was a compilation of results for data samples from all three subsets which had corresponding duration and temperature variables present within the dataset. 

\begin{figure}[H]
    \centering
    \includegraphics[width=\linewidth]{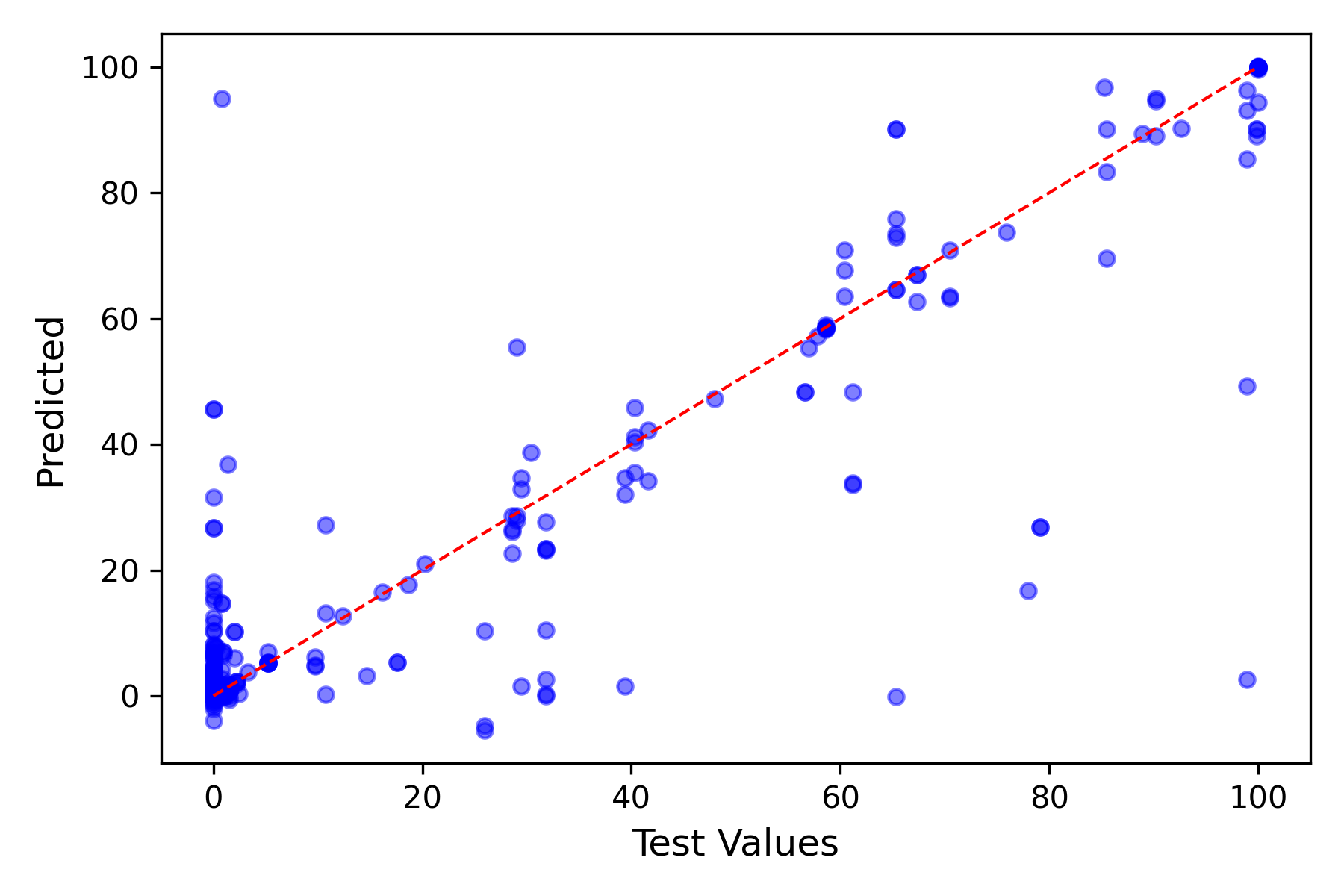}
    \captionsetup{font=small, labelfont=bf}
    \caption{Atomic percentage predictions of the elements Zn, Ti, Ni, Cu, Mn, Fe through the multi-targeted inverse regression ensemble model.}
    \label{fig:inverse_results}
\end{figure}

\section{Conclusion}
This study demonstrates the efficacy of AI/ML models in predicting and enhancing the corrosion resistance of aluminum alloys. The integration of various open-source datasets and ML workflows provides a scalable and efficient alternative to traditional methods.

We can see particularly the effect of proper model selection on the chosen accuracy metrics, and the evidence in this paper suggests that Gaussian Processes may be quite effective at modeling problems such as this. Through all versions of the forward model that were tested, we found that using an exact Gaussian process with a Matérn kernel yielded the best results, followed closely by a log-transformed Gaussian Process with a hybrid kernel function. It is, however, important to note that the complexity of a Gaussian process regression model is $O(n^3)$, so while for small datasets such as this, it may be effective, there is a substantial tradeoff when working with larger data. This can be circumvented by then using approximate or sparse Gaussian processes, or through the effective use of deep learning technologies, whose performance often scales with increased data availability. While errors in large values with the log-GPR were present, further modifications could include the use of other warping or transformation functions, in order to ensure both that the data is modeled accurately, and that errors are not exemplified when transforming back to the original space. In the interest of time, these methods were not explored, however have been used historically for materials property prediction problems.

Our forward models generally demonstrated improved performance at lower corrosion rates, though there was a noticeable decline in accuracy for higher corrosion rates. This may be due to the lack of adequate training data containing an abundant amount of larger corrosion measurement values. The exception to this, however, was the GPR model with the Matérn kernel, which predicted both lower and higher corrosion rate values with relatively higher efficiency. 

The inverse regression model also produced a highly effective framework through which we were able to predict alloy atomic composition values, given data on corrosion rates, environmental exposure conditions for the alloys, and main alloying element composition values. As is often true for ensemble models, we noticed that the presence of both a random forest regression as well as a gradient boosting one contributed to a higher overall predictive accuracy, being able to capture a variety of underlying trends in the data modeled, and combining each sub-model's strengths. 

For future work, this study can be improved by incorporating a greater diversity and volume of experimental data, in order to produce a more robust and generalizable analysis. Furthermore, the addition of computational data such as formation energies of materials chosen could provide an additional layer of insight into each alloy's mechanical properties as well as overall stability of their structures. Further research may also explore methods to quantify the uncertainty in these predictions to better assess model confidence, along with optimizing material compositions for enhancing corrosion protection and resistance.


\section{Acknowledgments}

This work was performed as part of The Washington Institute for STEM Entrepreneurship and Research Quantum Solutions Launchpad. The authors would like to thank Yaseen Ayoub for his contributions during the initial phase of the project. This manuscript has been co-authored by contractors of the U.S. Government under contract number DOE89233018CNR000004. Accordingly, the U.S. Government retains a non-exclusive, royalty-free license to publish or reproduce the published form of this contribution, or allow others to do so, for U.S. Government purposes.

\section{References}
\bibliographystyle{ieeetr}
\bibliography{ref}

\begin{thebibliography}{10}

\bibitem{reboul2011metallurgical}
M.~Reboul and B.~Baroux, ``Metallurgical aspects of corrosion resistance of
  aluminium alloys,'' {\em Materials and Corrosion}, vol.~62, no.~3,
  pp.~215--233, 2011.

\bibitem{lunder1997corrosion}
O.~Lunder, J.~Nordien, and K.~Nisancioglu, ``Corrosion resistance of cast mg-al
  alloys,'' {\em Corrosion Reviews}, vol.~15, no.~3-4, pp.~439--470, 1997.

\bibitem{esquivel2020corrosion}
J.~Esquivel and R.~Gupta, ``corrosion-resistant metastable al alloys: an
  overview of corrosion mechanisms,'' {\em Journal of The Electrochemical
  Society}, vol.~167, no.~8, p.~081504, 2020.

\bibitem{juan2021accelerating}
Y.~Juan, Y.~Dai, Y.~Yang, and J.~Zhang, ``Accelerating materials discovery
  using machine learning,'' {\em Journal of Materials Science \& Technology},
  vol.~79, pp.~178--190, 2021.

\bibitem{raabe2023accelerating}
D.~Raabe, J.~R. Mianroodi, and J.~Neugebauer, ``Accelerating the design of
  compositionally complex materials via physics-informed artificial
  intelligence,'' {\em Nature Computational Science}, vol.~3, no.~3,
  pp.~198--209, 2023.

\bibitem{li2020ai}
J.~Li, K.~Lim, H.~Yang, Z.~Ren, S.~Raghavan, P.-Y. Chen, T.~Buonassisi, and
  X.~Wang, ``Ai applications through the whole life cycle of material
  discovery,'' {\em Matter}, vol.~3, no.~2, pp.~393--432, 2020.

\bibitem{zhang2020bayesian}
Y.~Zhang, D.~W. Apley, and W.~Chen, ``Bayesian optimization for materials
  design with mixed quantitative and qualitative variables,'' {\em Scientific
  reports}, vol.~10, no.~1, p.~4924, 2020.

\bibitem{Jain2013}
A.~Jain, S.~P. Ong, G.~Hautier, W.~Chen, W.~D. Richards, S.~Dacek, S.~Cholia,
  D.~Gunter, D.~Skinner, G.~Ceder, and K.~A. Persson, ``{The Materials Project:
  A materials genome approach to accelerating materials innovation},'' {\em APL
  Materials}, vol.~1, no.~1, p.~011002, 2013.

\bibitem{merchant2023scaling}
A.~Merchant, S.~Batzner, S.~S. Schoenholz, M.~Aykol, G.~Cheon, and E.~D. Cubuk,
  ``Scaling deep learning for materials discovery,'' {\em Nature}, 2023.

\bibitem{ji2023corrosion}
Y.~Ji, X.~Fu, F.~Ding, Y.~Xu, Y.~He, M.~Ao, F.~Xiao, D.~Chen, P.~Dey, K.~Xiao,
  {\em et~al.}, ``Corrosion-resistant aluminum alloy design through machine
  learning combined with high-throughput calculations,'' {\em arXiv preprint
  arXiv:2312.15899}, 2023.

\bibitem{zeng2024machine}
C.~Zeng, A.~Neils, J.~Lesko, and N.~Post, ``Machine learning accelerated
  discovery of corrosion-resistant high-entropy alloys,'' {\em Computational
  Materials Science}, vol.~237, p.~112925, 2024.

\bibitem{smallfeng2019using}
S.~Feng, H.~Zhou, and H.~Dong, ``Using deep neural network with small dataset
  to predict material defects,'' {\em Materials \& Design}, vol.~162,
  pp.~300--310, 2019.

\bibitem{xu2023small}
P.~Xu, X.~Ji, M.~Li, and W.~Lu, ``Small data machine learning in materials
  science,'' {\em npj Computational Materials}, vol.~9, no.~1, p.~42, 2023.

\bibitem{ji2022randomsmall}
Y.~Ji, N.~Li, Z.~Cheng, X.~Fu, M.~Ao, M.~Li, X.~Sun, T.~Chowwanonthapunya,
  D.~Zhang, K.~Xiao, {\em et~al.}, ``Random forest incorporating ab-initio
  calculations for corrosion rate prediction with small sample al alloys
  data,'' {\em npj Materials Degradation}, vol.~6, no.~1, p.~83, 2022.

\bibitem{sasidhar2023enhancing}
K.~N. Sasidhar, N.~H. Siboni, J.~R. Mianroodi, M.~Rohwerder, J.~Neugebauer, and
  D.~Raabe, ``Enhancing corrosion-resistant alloy design through natural
  language processing and deep learning,'' {\em Science Advances}, vol.~9,
  no.~32, p.~eadg7992, 2023.

\bibitem{hu2024designing}
M.~Hu, Q.~Tan, R.~Knibbe, M.~Xu, G.~Liang, J.~Zhou, J.~Xu, B.~Jiang, X.~Li,
  M.~Ramajayam, {\em et~al.}, ``Designing unique and high-performance al alloys
  via machine learning: Mitigating data bias through active learning,'' {\em
  Computational Materials Science}, vol.~244, p.~113204, 2024.

\bibitem{tran2020multi}
A.~Tran, J.~Tranchida, T.~Wildey, and A.~P. Thompson, ``Multi-fidelity
  machine-learning with uncertainty quantification and bayesian optimization
  for materials design: Application to ternary random alloys,'' {\em The
  Journal of Chemical Physics}, vol.~153, no.~7, 2020.

\bibitem{ji2022random}
Y.~Ji, N.~Li, Z.~Cheng, X.~Fu, M.~Ao, M.~Li, X.~Sun, T.~Chowwanonthapunya,
  D.~Zhang, K.~Xiao, {\em et~al.}, ``Random forest incorporating ab-initio
  calculations for corrosion rate prediction with small sample al alloys
  data,'' {\em npj Materials Degradation}, vol.~6, no.~1, p.~83, 2022.

\bibitem{harsimran2021overview}
S.~Harsimran, K.~Santosh, and K.~Rakesh, ``Overview of corrosion and its
  control: A critical review,'' {\em Proc. Eng. Sci}, vol.~3, no.~1,
  pp.~13--24, 2021.

\bibitem{riggs1967temperature}
O.~L. Riggs~Jr and R.~M. Hurd, ``Temperature coefficient of corrosion
  inhibition,'' {\em Corrosion}, vol.~23, no.~8, pp.~252--260, 1967.

\bibitem{Ricker1997}
R.~Ricker, ``{CORR-DATA},'' 1997.

\bibitem{MG-ATRENS2020989}
A.~Atrens, Z.~Shi, S.~U. Mehreen, S.~Johnston, G.-L. Song, X.~Chen, and F.~Pan,
  ``Review of mg alloy corrosion rates,'' {\em Journal of Magnesium and
  Alloys}, vol.~8, no.~4, pp.~989--998, 2020.

\bibitem{Knovel_Aluminum_Database}
Knovel, ``Knovel aluminum database.''
  \url{https://app.knovel.com/kn/resources/kpAAD00001/toc}.

\bibitem{Vargel2020}
C.~Vargel, J.-M. Germain, and H.~Dunlop, {\em Corrosion of Aluminium}.
\newblock Elsevier, 2nd~ed., 2020.

\bibitem{osti_1069242}
R.~W. and RA~Bonewitz, ``Catalog information on the performance of aluminum in
  sea water,'' tech. rep., Pacific Northwest National Lab. (PNNL), Richland, WA
  (United States), 04 1978.

\bibitem{MatWeb2025}
{MatWeb, LLC}, ``Online materials database.'' \url{https://www.matweb.com/},
  2025.

\bibitem{Aluminum2025}
{The Aluminum Association}, ``Aluminum alloy information.''
  \url{https://www.aluminum.org/}, 2025.

\bibitem{Copper2025}
{Copper Development Association}, ``Copper alloys database.''
  \url{https://www.copper.org/}, 2025.
\newblock Accessed: 2025-03-01.

\bibitem{sklearn_api}
L.~Buitinck, G.~Louppe, M.~Blondel, F.~Pedregosa, A.~Mueller, O.~Grisel,
  V.~Niculae, P.~Prettenhofer, A.~Gramfort, J.~Grobler, R.~Layton,
  J.~VanderPlas, A.~Joly, B.~Holt, and G.~Varoquaux, ``{API} design for machine
  learning software: experiences from the scikit-learn project,'' in {\em ECML
  PKDD Workshop: Languages for Data Mining and Machine Learning}, pp.~108--122,
  2013.

\bibitem{ISL2013}
G.~James, D.~Witten, T.~Hastie, and R.~Tibshirani, {\em An Introduction to
  Statistical Learning}, vol.~103 of {\em Springer Texts in Statistics}.
\newblock Springer, 2013.

\bibitem{takamoto2022towards}
S.~Takamoto, C.~Shinagawa, D.~Motoki, K.~Nakago, W.~Li, I.~Kurata, T.~Watanabe,
  Y.~Yayama, H.~Iriguchi, Y.~Asano, {\em et~al.}, ``Towards universal neural
  network potential for material discovery applicable to arbitrary combination
  of 45 elements,'' {\em Nature Communications}, vol.~13, no.~1, p.~2991, 2022.

\bibitem{witman2023defect}
M.~D. Witman, A.~Goyal, T.~Ogitsu, A.~H. McDaniel, and S.~Lany, ``Defect graph
  neural networks for materials discovery in high-temperature clean-energy
  applications,'' {\em Nature Computational Science}, vol.~3, no.~8,
  pp.~675--686, 2023.

\bibitem{DNN_FENG2019300}
S.~Feng, H.~Zhou, and H.~Dong, ``Using deep neural network with small dataset
  to predict material defects,'' {\em Materials \& Design}, vol.~162,
  pp.~300--310, 2019.

\bibitem{Approximator}
K.~Hornik, M.~Stinchcombe, and H.~White, ``Multilayer feedforward networks are
  universal approximators,'' {\em Neural Networks}, vol.~2, no.~5,
  pp.~359--366, 1989.

\bibitem{Goodfellow-et-al-2016}
I.~Goodfellow, Y.~Bengio, and A.~Courville, {\em Deep Learning}.
\newblock MIT Press, 2016.
\newblock \url{http://www.deeplearningbook.org}.

\bibitem{pytorch}
A.~Paszke, S.~Gross, F.~Massa, A.~Lerer, J.~Bradbury, G.~Chanan, T.~Killeen,
  Z.~Lin, N.~Gimelshein, L.~Antiga, A.~Desmaison, A.~Kopf, E.~Yang, Z.~DeVito,
  M.~Raison, A.~Tejani, S.~Chilamkurthy, B.~Steiner, L.~Fang, J.~Bai, and
  S.~Chintala, ``Pytorch: An imperative style, high-performance deep learning
  library,'' in {\em Advances in Neural Information Processing Systems 32},
  pp.~8024--8035, Curran Associates, Inc., 2019.

\bibitem{Huberhastie2009elements}
T.~Hastie, R.~Tibshirani, J.~Friedman, {\em et~al.}, ``The elements of
  statistical learning,'' 2009.

\bibitem{GP_wang2023}
J.~Wang, ``An intuitive tutorial to gaussian process regression,'' {\em
  Computing in Science \& Engineering}, vol.~25, no.~4, pp.~4--11, 2023.

\bibitem{gaussian_williams2006}
C.~K. Williams and C.~E. Rasmussen, {\em Gaussian processes for machine
  learning}, vol.~2.
\newblock MIT press Cambridge, MA, 2006.

\bibitem{Materngenton2001classes}
M.~G. Genton, ``Classes of kernels for machine learning: a statistics
  perspective,'' {\em Journal of machine learning research}, vol.~2, no.~Dec,
  pp.~299--312, 2001.

\bibitem{stein1999interpolation}
M.~L. Stein, {\em Interpolation of spatial data: some theory for kriging}.
\newblock Springer Science \& Business Media, 1999.

\bibitem{additive_hastie2017generalized}
T.~J. Hastie, ``Generalized additive models,'' {\em Statistical models in S},
  pp.~249--307, 2017.

\bibitem{duvenaud2014automatic}
D.~Duvenaud, {\em Automatic model construction with Gaussian processes}.
\newblock PhD thesis, University of Cambridge, 2014.

\bibitem{GPsnelson2003warped}
E.~Snelson, Z.~Ghahramani, and C.~Rasmussen, ``Warped gaussian processes,''
  {\em Advances in neural information processing systems}, vol.~16, 2003.

\bibitem{GPswiler2020survey}
L.~P. Swiler, M.~Gulian, A.~L. Frankel, C.~Safta, and J.~D. Jakeman, ``A survey
  of constrained gaussian process regression: Approaches and implementation
  challenges,'' {\em Journal of Machine Learning for Modeling and Computing},
  vol.~1, no.~2, 2020.

\bibitem{gpytorch}
J.~R. Gardner, G.~Pleiss, D.~Bindel, K.~Q. Weinberger, and A.~G. Wilson,
  ``Gpytorch: Blackbox matrix-matrix gaussian process inference with gpu
  acceleration,'' in {\em Advances in Neural Information Processing Systems},
  2018.

\bibitem{xgboost}
T.~Chen and C.~Guestrin, ``{XGBoost}: A scalable tree boosting system,'' in
  {\em Proceedings of the 22nd ACM SIGKDD International Conference on Knowledge
  Discovery and Data Mining}, KDD '16, (New York, NY, USA), pp.~785--794, ACM,
  2016.

\bibitem{kritzer2004corrosion}
P.~Kritzer, ``Corrosion in high-temperature and supercritical water and aqueous
  solutions: a review,'' {\em The Journal of Supercritical Fluids}, vol.~29,
  no.~1-2, pp.~1--29, 2004.

\bibitem{bhat2024inverse}
N.~Bhat, A.~S. Barnard, and N.~Birbilis, ``Inverse design of aluminium alloys
  using multi-targeted regression,'' {\em Journal of Materials Science},
  vol.~59, no.~4, pp.~1448--1463, 2024.

\end{thebibliography}


\end{document}